\documentclass[preprint,3p,twocolumn,12pt]{elsarticle}

\usepackage{hyperref}
\usepackage{amssymb}
\usepackage{amsmath}

\graphicspath{{./Figures/}{./}}
\journal{Nuclear Physics B}

\begin{document}

\begin{frontmatter}



\title{Experimental study of the acoustic frequency-up conversion effect by nonlinear thin plates} 

 \author[label1]{Alexis Mousseau}
 \affiliation[label1]{organization={Laboratoire d'Acoustique de l'Université du Mans (LAUM), UMR 6613, Institut d'Acoustique - Graduate School (IA-GS), CNRS, Le Mans Université, France}}

 \author[label1]{Soizic Terrien} 

  \author[label1]{Vincent Tournat}


\begin{abstract}
Classical methods of sound absorption present fundamental limits that can be overcome by using nonlinear effects. Thin clamped plates have been identified as strongly nonlinear elements, capable of transferring the acoustic power of an incident air-borne wave towards higher frequencies. Here, we experimentally show that these plates exhibit different vibrational nonlinear behaviors depending on the amplitude and frequency of the excitation signal. The lowest excitation levels achieved lead to harmonic generation in a weakly nonlinear regime, while higher levels produce quasi-periodic and chaotic regimes. Since these nonlinear vibration regimes govern the acoustic frequency-up conversion process, we investigate the influence of relevant physical and geometrical parameters on the emergence of these nonlinear regimes. A parametric study on plates of different thicknesses reveals that the frequency-up conversion effect is mostly guided by the resonance of the plate at its first eigenfrequency, which depends not only on its thickness but also on a static tension introduced by the clamping. Finally, a design proposition involving multiple plates with different properties is presented in order to reach a broadband frequency-up conversion.\end{abstract}




\end{frontmatter}



Controlling the propagation of sound with metamaterials is crucial for noise control innovative strategies \cite{yang_sound_2017,deymier_acoustic_2013,craster_acoustic_2024}. Early solutions to absorb airborne sound waves relied on porous materials \cite{allard_propagation_2009, bolton_elastic_1997,cao_porous_2018}, which are effective above a certain frequency determined by their properties including their thickness. As a result, in practice, large thicknesses of a few dozen of cm are typically required to absorb frequencies below 500 Hz. Therefore, in recent years, there has been a growing interest in the development of acoustic metamaterials \cite{cummer_controlling_2016,krushynska_emerging_2023}, which mostly rely on a combination of repeating patterns and resonances to achieve unprecedented sound manipulation  \cite{li_double-negative_2004,norris_acoustic_2008,jimenez_metadiffusers_2017,jimenez_rainbow-trapping_2017,mallejac_zero-phase_2019,mallejac_doping_2020}. The majority of the metamaterials developed to absorb sound waves are typically passive and linear, which imposes fundamental absorption limitations, namely a trade-off between absorption bandwidth and geometrical configurations, even if they can be found efficient at sub-wavelength thicknesses \cite{yang_optimal_2017,yang_integration_2018}. This work proposes an approach for controlling acoustic waves that relies on passive nonlinear effects to overcome the fundamental limitations of linear systems.

In airborne acoustics, nonlinear phenomena are generally difficult to observe and even more challenging to exploit. A well-known instance of nonlinear behavior is the propagation of high-amplitude sound waves, which can lead to the formation of shock waves \cite{myers_effects_2012,hamilton_nonlinear_2024} due to cumulative propagative effects.  In contrast, local nonlinear phenomena, which we aim at using here, are less common and harder to achieve. An example in the neighboring field of water acoustics is the nonlinear response of air bubbles excited near their fundamental resonance frequencies \cite{leighton_4_1994}. However, such local nonlinear elements as air bubbles in water do not have their equivalent in airborne acoustics. 

This work stems from the observation of highly nonlinear elements, specifically thin clamped plates, which exhibit extremely nonlinear vibrations in reaction to acoustic excitation. These elements were initially used in the conception of a membrane-based acoustic metamaterial \cite{mallejac_zero-phase_2019}, where the nonlinear effects were considered a hindrance. In contrast, we aim here to investigate and take advantage of these highly nonlinear and local effects. When subject to sinusoidal acoustic excitation, these elements exhibit a range of dynamic behaviors far more complex than the original excitation. Notably, these dynamical regimes lead to observable changes in the radiated sound, which becomes audibly enriched with harmonics or, in the case of nonperiodic vibrations, takes on a "kazoo-like" quality. Both of these effects result in an enrichment of the acoustic power to higher frequencies than the original excitation. 

Previous studies investigated plates of larger dimensions, both circular \cite{givois_experimental_2020,touze_idiophones_2016,chaigne_nonlinear_2005} and rectangular in shape \cite{amabili_nonlinear_2018,amabili_nonlinear_2004,amabili_theory_2006}, under different boundary conditions, providing insight into the dynamics of these objects. These studies highlighted the dependency of the eigenfrequencies on the amplitude of forcing, a phenomenon referred to as "hardening" or "softening" of the so-called nonlinear modes. The emergence of quasiperiodic vibrations has also been observed and explained by a phenomenon referred to as internal resonances, which occur when there is a specific algebraic relationship between the eigenfrequencies of two modes \cite{thomas_asymmetric_2003}. This typically arises when the frequency of one mode is approximately an integer multiple of another, such as $f_2 \approx n f_1$, with $n$ an integer. These phenomena are linked to the emergence of different nonlinear dynamical regimes, which range from weakly nonlinear regimes characterized by a generation of harmonics to a quasiperiodic regime when there is an internal resonance and a chaotic \cite{awrejcewicz_spatio-temporal_2002} (or turbulent \cite{during_weak_2006}) regime when the forcing amplitude is particularly strong. The emergence of these regimes is dictated by the vibrational modes, which depend on several factors such as the geometrical parameters and boundary conditions of the plate.

This study aims to investigate experimentally the behavior of thin plates under acoustic excitation, in order to characterize and maximize the observed acoustic frequency-up conversion. Section II presents the experimental setup used to collect information on the acoustic field and the vibration of the plate. Section III shows the investigation of the vibration under different frequencies and amplitudes of excitation, in addition to a parametric study on the role of thickness in the dynamics of the plates. Section VI demonstrates the impact of the vibration on the acoustic frequency-up conversion. Finally, Section V  explores the efficiency of an arrangement of plates with different thicknesses and sizes in panels, aiming to observe broadband frequency-up conversion.

\begin{figure*}[t]
    \centering
    \includegraphics[width=0.98\linewidth]{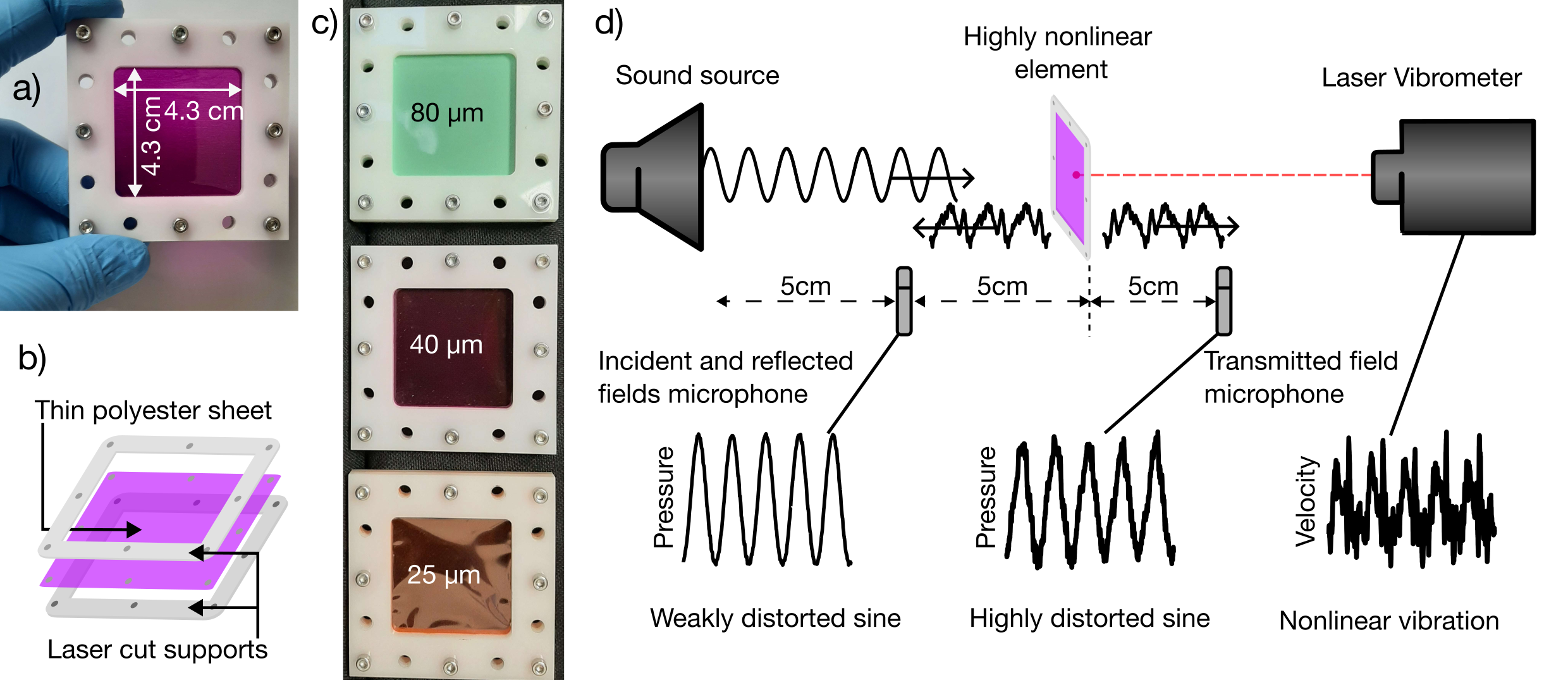}
    \caption{Samples and experimental configuration. a) Photography of the nonlinear plate with the clamping frame, b) Schematics of the clamping of the polyester sheet in the supports to build the nonlinear element, c) Photography of the three tested polyester sheets of different thicknesses, d) Schematics of the experimental setup with typical signals measured at the three different points of measurement.}
    \label{fig:Schema}
\end{figure*}

\section{Problem and experimental setup}



The nonlinear system investigated in this study consists of a thin sheet of polyester clamped in a square rigid frame, as depicted in Figure \ref{fig:Schema}a-c. The clamping is secured using bolts, tightened with a torque wrench to ensure that it is uniform, allowing a better reproducibility of the experimental results. Once clamped, the vibrating portion of the object forms a square thin plate with dimensions of 4.3$\times$4.3 cm. The color of each sheet indicates its thickness: green corresponds to 80$\mu$m, purple to 40$\mu$m, and orange to 25$\mu$m. 

The experimental setup, sketched in Figure \ref{fig:Schema}, consists of a loudspeaker positioned 10 cm away from this nonlinear scatterer. A first microphone is placed between the loudspeaker and the object to capture both the acoustic field emitted by the sound source and the reflection from the nonlinear scatterer. The second microphone is positioned 5 cm behind the object to capture the transmitted acoustic signal. Additionally, a laser vibrometer is focused just off-center of the vibrating plate, in order to avoid as much as possible vibration nodal lines of the first few modes, and captures the velocity of the plate at this point. These three measurement points enable us to characterize some of the vibrational behavior of the scatterer, and then relate it to its effects on the acoustic field.

To characterize the nonlinear response of this scatterer, the excitation signal is chosen to map the frequency/amplitude parametric space. Two choices can then be made, either keeping the amplitude constant while varying the frequency, by using a swept sine, or maintaining a constant frequency while the amplitude changes. 

The first possibility presents a technical issue linked to the method of excitation. Loudspeakers, by their design, typically have their own resonances and frequency responses, which complicates the process of mapping the amplitude parameter space, as a constant amplitude in Volts given to the loudspeaker would lead to a pressure level varying with frequency. For this reason, the chosen signal is swept in amplitude rather than frequency over time. Another issue pertains to the amplitude sweep, as a continuous change of frequency or amplitude would make it difficult to differentiate between transient and steady-state regimes. 
For these reasons, we opt for the second approach, using an amplitude-modulated sine wave with a fixed frequency. The envelope of this sine wave is described by a sum of time-shifted hyperbolic tangents: 
 
\begin{multline} \label{eq:envelope}   
    f(t) =\sum_{n=1}^N[G(n) -G(n-1)]\\
    [1+\tanh{(S(t-(n-1)T_{step}-dl}]/2,
\end{multline}

\noindent
where $N$ is the number of steps in the signal, $S$ is a smoothness parameter (inverse of a transition characteristic time), $t$ is the time vector, $T_{step}$ is the time allocated to each amplitude step, $dl$  is the transition time between two amplitude steps, $G(n)$ is a vector of $N+1$ elements, where the first value is zero and  each subsequent value corresponds to the amplitude of each step. Specifically $G(n)=[0,A,A\cdot st,...,A\cdot st^{n-1}]$, with $st$ the multiplicative factor between steps.

The chosen signal, displayed in Fig.~\ref{fig:Schema}.a, consists of 10 steps lasting 5 seconds, each step represents a 2 dB increment in pressure level, which corresponds to the parameters $N=10$, $S=2$, $T_{step}=5$, $dl=0.2$, and $st=10^{0.1}$. This signal avoids the problem of varying excitation amplitude by using steps instead of a continuous sweep in amplitude, but it is still sensitive to the frequency response of the loudspeaker. To address this issue, a distinct voltage value is assigned to each frequency, ensuring a consistent sound pressure level across all the frequencies tested.

\section{Vibrational behavior of thin plates under acoustic excitation}

The experimental setup described above is used to study the dynamics of our nonlinear plates under acoustic excitation. The first part of the investigation focuses on demonstrating frequency-up conversion induced by the vibration of the plate, for a single excitation frequency. In the second part, we explore the vibrational behavior within the amplitude/frequency parameter space. Lastly, we examine the effect of the different dynamic regimes on the surrounding acoustic field. 
\subsection{Vibrational response to a single frequency excitation at different amplitudes}
\begin{figure*}[ht]
    \centering
    \includegraphics[width=0.98\linewidth]{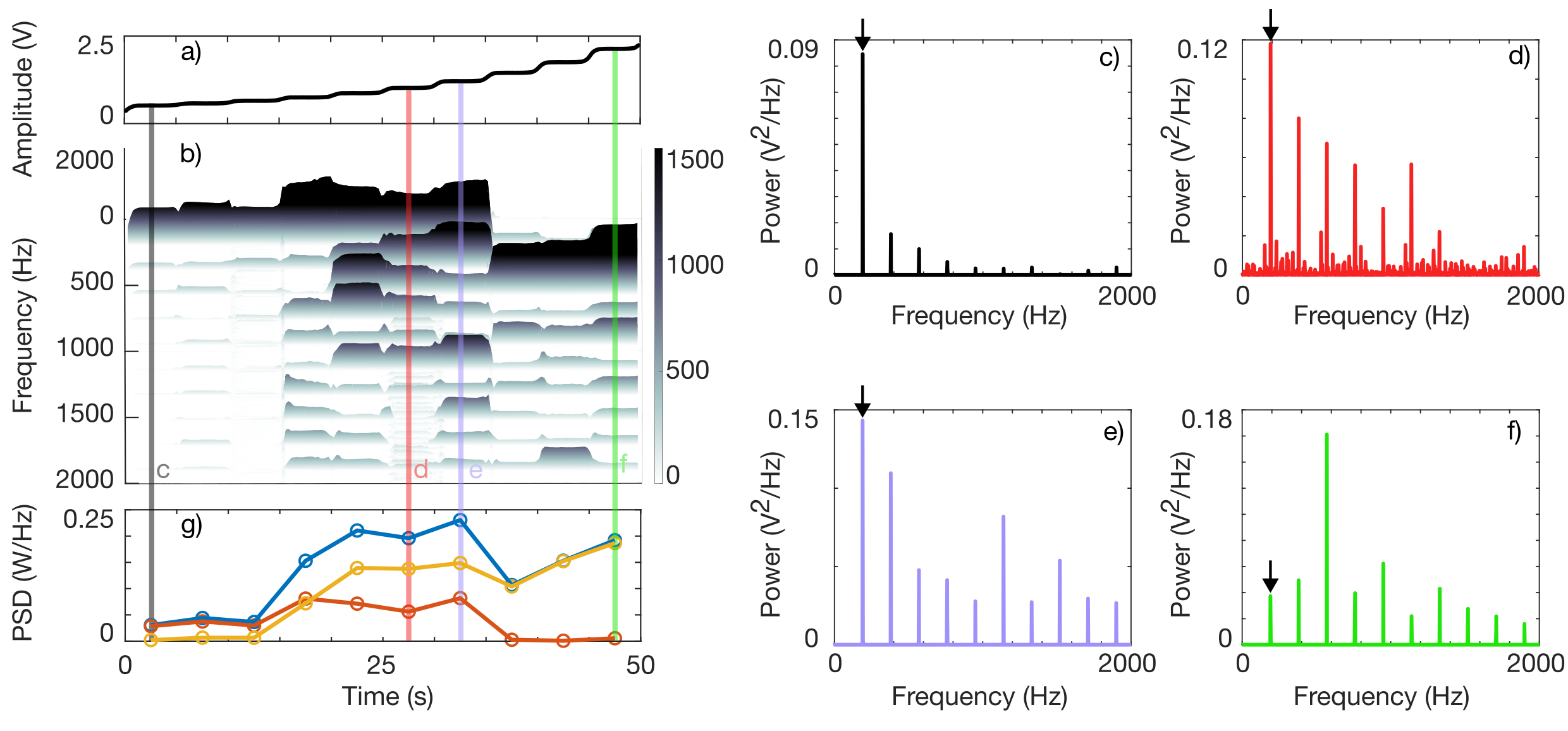}
    \caption{Vibrational behavior of the 40 $\mu m$ plate as observed experimentally when acoustically excited with an amplitude modulated sine at 190 Hz. a) Amplitude of the input sinusoidal signal. b) Spectrogram of the vibration measured at a point near the center of a 40 $\mu m$ plate, in response to an acoustic excitation at 190 Hz of rising amplitude between 77 and 95 dB. Power spectrum of the out of plane velocity signal (from the vibrometer) between : c) 1 and 3 seconds (77 dB incident), d) 26 and 29 seconds (87 dB incident), e) 31 and 33 seconds (89 dB incident), f) 46 and 49 seconds (95 dB incident). g) Power spectral density between 0 and 20 kHz (in blue), in a band of 10 Hz around the frequency of excitation (in red), and above the frequency of excitation (in yellow).}
    \label{fig:VibRegimes}
\end{figure*}

The first experiment consists of monitoring the vibration of the 40 $\mu$m plate in response to a sine excitation at 190 Hz, for varied excitation levels from 77 dB to 95 dB. The experimental results are displayed in Fig.~\ref{fig:VibRegimes}.

\noindent
Figure \ref{fig:VibRegimes}.a displays the amplitude modulation of the excitation signal as described in the equation \eqref{eq:envelope}.
The spectrogram of the velocity near the center of the plate, shown in Fig.~\ref{fig:VibRegimes}.b. reveals a complex dynamics that can be apprehended by analyzing the plate vibration at specific time intervals individually. At the lowest excitation level (77 dB), harmonic generation can already be observed, as shown on the spectrum displayed in Fig.~\ref{fig:VibRegimes}.c. Although this does not significantly contribute to the total vibration of the plate, which mostly occurs around the excitation frequency, it is noteworthy that such a low excitation level can still trigger nonlinear frequency-up conversion. As the excitation level increases to 87 dB, we observe the emergence of broadband non-periodic vibrations in Fig.~\ref{fig:VibRegimes}.d, appearing in the spectrum as multiple peaks located at frequencies that are not integer multiples of the excitation frequency. These peaks are approximately one order of magnitude lower in power than the fundamental frequency. The harmonics are also more predominant, showing significant level up to the sixth harmonics. When the excitation level is increased to 89 dB, the dynamic regime changes and the system enters a strongly harmonic-distorted regime, as shown in Fig.~\ref{fig:VibRegimes}.e, where the second harmonics has a power nearly equal to that of the fundamental frequency.
From 91 dB to 95 dB, the energy becomes predominantly located on the third harmonic of the excitation, as can be observed on the spectrum in Fig.~\ref{fig:VibRegimes}.f, even surpassing the fundamental frequency energy.

The spectrogram and spectra illustrate the various regimes that arise when the excitation level is changed, highlighting the complexity and nonlinearity of the plate dynamic behavior. One possible interpretation of the change in dynamics when the excitation level is increased is that, from 0 to 15 seconds (77 dB to 81dB incident), the system is excited far from its first resonance frequency, and responds mostly linearly to the excitation, with a maximum of energy at the frequency of excitation and low-amplitude harmonics, resulting from the relatively large vibration amplitude compared to the plate's thickness. Between 15 and 35 seconds (83dB to 89dB incident), due to the resonance frequency depending on the excitation level, the system enters a resonating regime, which leads to the emergence of higher harmonics at amplitudes of the same order of magnitude as the fundamental due to the higher amplitude of displacement. After 35 seconds (91 dB incident), the maximum peak in the frequency spectrum, which we denote as the dominant frequency throughout this article,  switches to the third harmonics, resembling the phenomenon of internal resonance observed in other plate-like systems \cite{thomas_asymmetric_2003,chaigne_nonlinear_2005,touze_asymetric_2002}. At these specific frequencies and amplitudes, it is more favorable for the plate to vibrate in a higher mode that is a multiple of the current excitation frequency, in this case, 190 Hz, effectively concentrating the power on the third harmonics of the excitation.

The Figure \ref{fig:VibRegimes}.g supports this hypothesis regarding the dynamics, by showing the changes in power spectral density (PSD) of the velocity signal over time. The blue line represents the total PSD between 0 and 20 kHz, the red one, in a band of 20 Hz around the excitation frequency, here from 180 to 200 Hz, and the yellow line is the PSD from 200 Hz to 20 kHz. The first three data points show that  most of the power is located in the fundamental frequency. When the system starts resonating at 15 seconds (83 dB incident), the power is almost equally distributed between the fundamental frequency and the higher frequencies. From 35 seconds (91 dB incident) to the end, most of the power is located above the excitation frequency, as shown on the spectrogram where the third harmonics is the dominant frequency component.

The response to a single frequency acoustic excitation provides valuable insight into the complex dynamics of the nonlinear scatterer. It is observed that this resonant system exhibits plate-like behaviors, notably a generation of harmonics depending on the amplitude of excitation, and a phenomenon reminiscent of internal resonance, where the excitation at $f_{exc}$ leads to a majority of the vibrational energy being transferred at $3f_{exc}$. The experiment is then repeated with different excitation frequencies ranging from 100 to 500 Hz, in 5 Hz increments.

\subsection{Vibration mapping in the amplitude/frequency space}
\begin{figure*}[ht!]
    \centering  \includegraphics[width=0.98\linewidth]{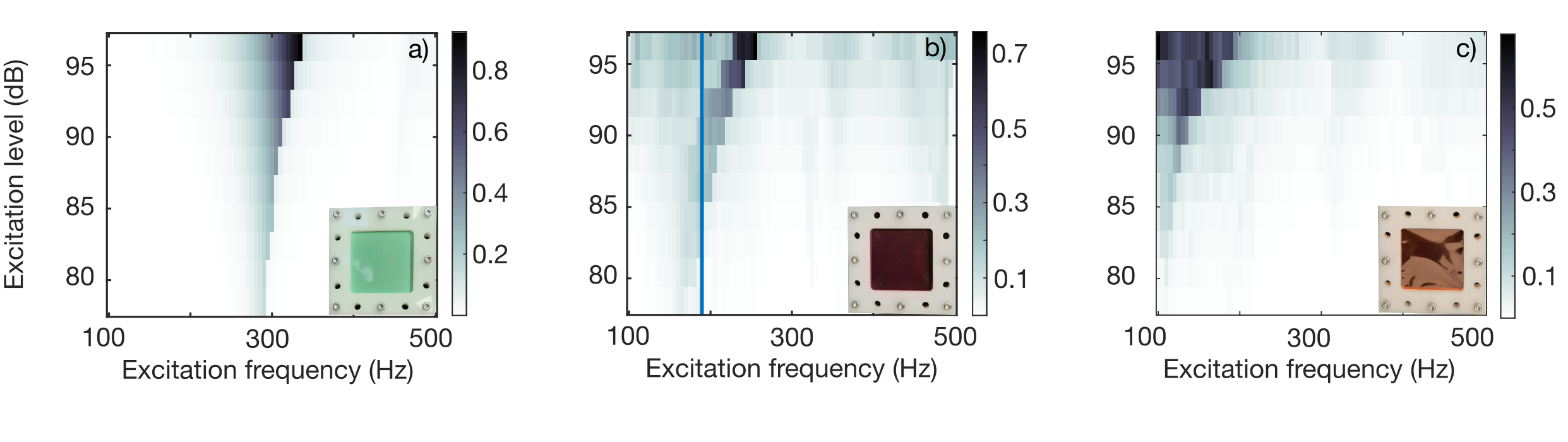}
    \caption{ Power Spectral Density map of the velocity measured near the center of the plate. a) For the 80$\mu$m plate. b) For the 40$\mu$m plate. c) For the 25$\mu$m plate.}
    \label{fig:ParamVib}
\end{figure*}

The maps in Fig.~\ref{fig:ParamVib} are a generalization of the results shown in blue in Figure \ref{fig:VibRegimes}.g, displaying the total vibrational power for each frequency and amplitude. Three different thicknesses have been tested, 80 $\mu$m , 40 $\mu$m  and 25 $\mu$m respectively displayed in Fig.~\ref{fig:ParamVib}.a, Fig.~\ref{fig:ParamVib}.b and Fig.~\ref{fig:ParamVib}.c. For each thickness, we observe a tongue of maxima, in a  darker shade of gray, which shifts towards the higher frequencies as the excitation level increases. Figure \ref{fig:ParamVib}.b unveils the dynamics that is observed with the 190 Hz excitation, further supporting the interpretation of the previous results. Let us call $f_{exc}$ the constant frequency of forcing, 190 Hz in this case, and $f_1(A)$ the frequency of the first mode of the plate, that depends on the amplitude of vibration. 

From 77 to 79 dB of excitation, $f_{exc}>f_1$, the plate is out of resonance, because the resonance frequency is lower than the excitation frequency. This is illustrated by the blue line in Fig \ref{fig:ParamVib}.b, which represents the 190 Hz experimental data, and that is located higher in frequency than the tongue for the lowest amplitudes. As the level of excitation increases from 81 to 89 dB, $f_{exc} \approx f_1$ by shifting up the resonance of the system, we observe in Fig.~\ref{fig:VibRegimes}.b) a highly distorted regime with a signal power mainly located in the fundamental and the first few harmonics. This is visible in Fig.~\ref{fig:ParamVib}.b, as the blue line coincides with the tongue of maxima between 81 and 89 dB. When the excitation level exceeds 89 dB, the plate resonance frequency shifts further up, $f_{exc}<f_1$, and as a result, the amplitude of the response at the fundamental frequency diminishes abruptly while the third harmonics of the excitation becomes the major component of the vibration. This sudden change in dynamical regime could be tied to an internal resonance, a commensurability between the frequency of excitation and the frequency of a higher mode, in this case $f_2(A)\approx 3f_{exc} $, with $f_2(A)$ the resonance frequency of a higher-order mode of the plate. Interestingly, there is no resonance tongue at higher frequencies that would correspond to higher modes, which we would expect from the previous observations on plates \cite{amabili_nonlinear_2004,amabili_nonlinear_2018,amabili_theory_2006,touze_idiophones_2016}. The lack of observable higher modes can be explained by the nature of the acoustic pressure forcing applied to the plate. 

\begin{figure}
    \centering
    \includegraphics[width=0.99\linewidth]{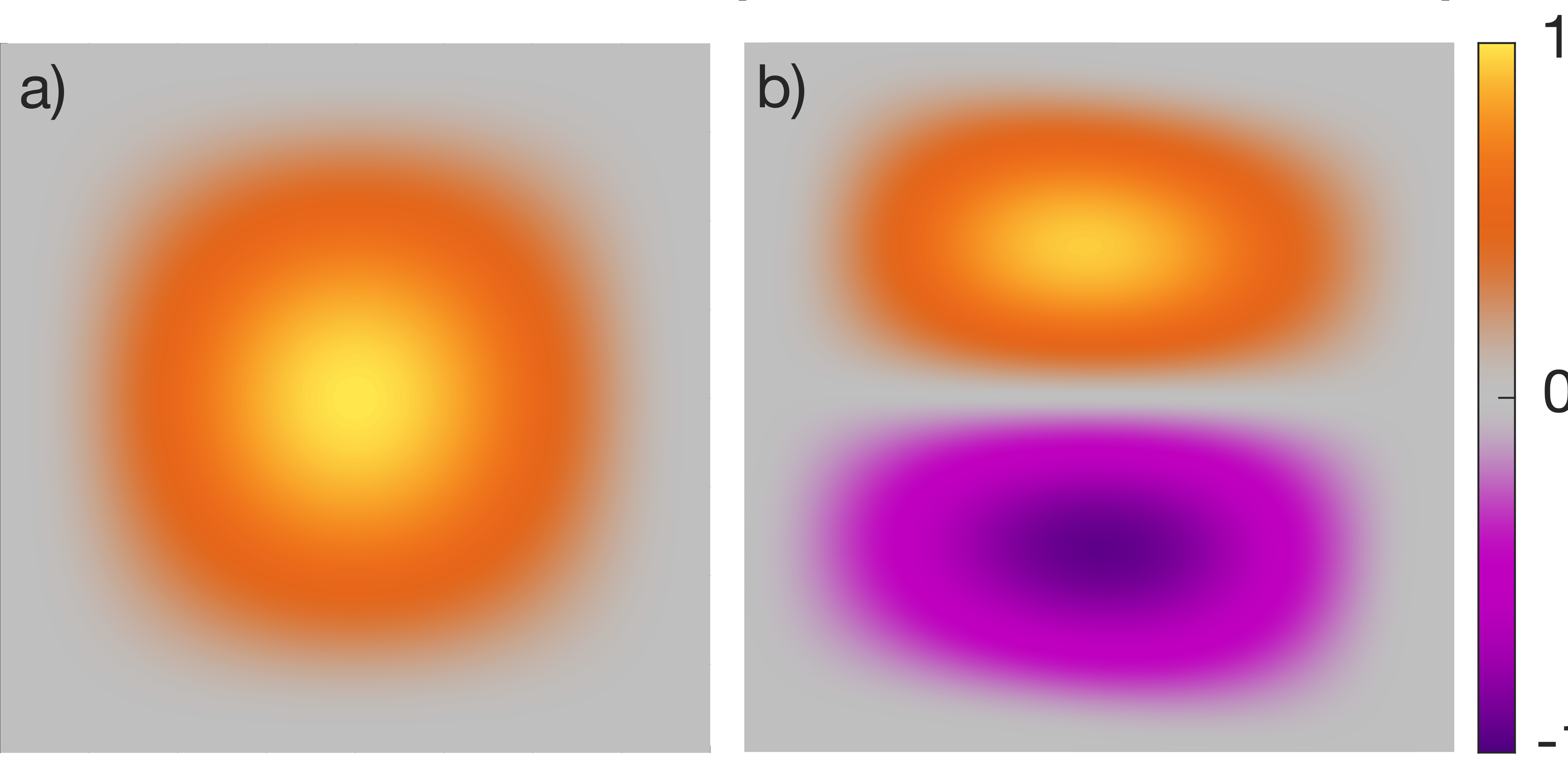}
    \caption{Normalized mode shapes of a square clamped plate. These mode shapes are computed using the finite-difference based algorithm MAGPIE \cite{hamilton_magpie_2024}. a) First "monopolar" mode shape b) Second "dipolar" mode shape. The acoustic excitation, a uniform pressure on the plate, couples efficiently with the monopolar mode, but cannot excite the dipolar mode for symmetry reason.}
    \label{fig:2modesnum}
\end{figure}

The acoustic excitation, acts as a uniform force applied on one side of the plate. An analytical modal analysis demonstrates \cite{ducceschi_modal_2015} that this type of forcing couples strongly with the first natural mode, depicted in Fig.~\ref{fig:2modesnum}.a, and weakly with any type of multipolar modes such as the second one, depicted in Fig.~\ref{fig:2modesnum}.b. This preferential coupling of the acoustic excitation to the first mode explains the single tongue of resonance observed.

\subsection{Influence of plate thickness on the dynamics}

When comparing the vibrational behavior of the three different thickness plates in Fig.~\ref{fig:ParamVib}, a direct relationship between thickness and resonance frequency is visible. According to the linear Kirchhoff-Love model \cite{love_small_1888}, independently of boundary conditions \cite{hamilton_magpie_2024,li_exact_2009,li_simple_2009}, the frequency of the first mode should increase linearly with the thickness, 
\begin{equation}\label{eq:anfreq}
    \omega_1=h\sqrt{\dfrac{E}{12\rho (1-\nu^2) }}\sqrt\lambda,
\end{equation}
\noindent
where $E$ is the Young's modulus, $h$ the thickness of the plate, $\rho$ the volumetric density, $\nu$ the Poisson ratio of the plate and $\lambda$ a value depending on the size and boundary condition of the plate. This equation implies that the first resonance frequency of the 80 $\mu m$ plate should be twice that of the 40 $\mu m$ plate if the studied scatterers were behaving purely as thin plates. However, the experimental results do not exhibit this linear relationship between thickness and first resonance frequency. A potential explanation of this discrepancy could be the static tension introduced by the clamping. The latter tends to shift the resonance frequencies higher than the expected results derived from the Kirchhoff-Love model.

\begin{figure*}[ht!]
    \centering
    \includegraphics[width=0.98\linewidth]{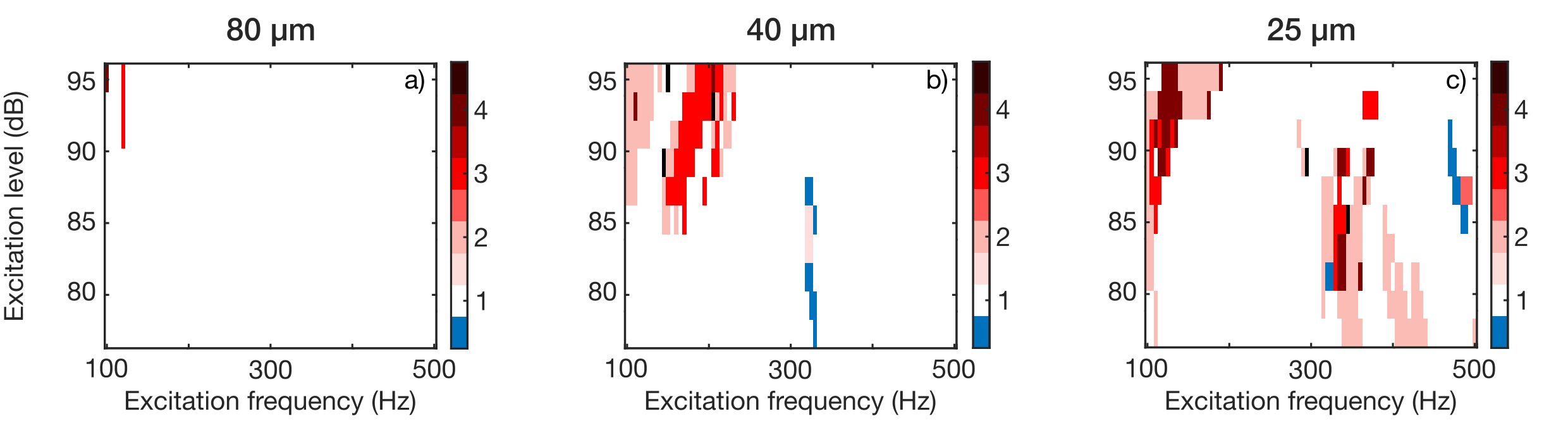}
    \caption{Dominant harmonics in the vibration signal plotted against the level and frequency of excitation for three plate thicknesses, a) 80 $\mu m$, b) 40 $\mu m$ and c) 25 $\mu m$. The colorbars indicate in which part of the spectrum the dominant frequency component lies, 1 (white) being the fundamental component, 2 (dark pink) the second harmonic component and intermediate colors standing for subharmonic 1/2 (blue) and 3/2, 5/2...}
    \label{fig:Paramharm}
\end{figure*}

In Figure \ref{fig:Paramharm}, the dominant vibration frequency is plotted as a function of excitation level and frequency. In these maps, white corresponds to regimes where the excitation frequency (the fundamental component) is the predominant vibration component, red corresponds to regimes where the major part of the energy in the vibrational signal is transferred to a higher frequency and blue to a lower one. Figure \ref{fig:Paramharm}.a displays the dominant harmonics map for the 80 $\mu$m plate. The predominant vibration frequency always matches the excitation frequency, with the exception of two values of the excitation frequency at high amplitudes. In contrast, the behavior of the 40 $\mu$m plate displayed in Fig.~\ref{fig:Paramharm}.b is very different. A zone of dominant third harmonic can be observed from 150 to 210 Hz, which corresponds to the regime shown previously in Fig.~\ref{fig:VibRegimes}.g. The dominant harmonic map of the 25 $\mu$m thick plate, depicted in Fig.~\ref{fig:Paramharm}.c, shows the most complex behavior of the three plates of different thicknesses tested. The maximum of vibration amplitude corresponds to the dominant fundamental regime up to 93 dB where the second harmonic becomes the dominant component of the vibration.
Figure \ref{fig:Paramharm} demonstrates that the system dynamics is strongly influenced by the plate thickness. This parameter determines the frequency of the first mode, which in turn defines the tongue of maximum velocity in the vibrational signal. The parametric study reveals that the thickest plates tend to behave more linearly to a given acoustic excitation. In contrast, the thinner plates display complex nonlinear behaviors around the resonance, notably dynamic regimes where the vibrational power is located outside of the excitation frequency, likely linked to internal resonance phenomenon. In the following section, we describe how the vibrational behavior affects the surrounding acoustic field and consequently the acoustic scattering properties of our nonlinear thin plates.

\begin{figure*}[ht!]
    \centering
    \includegraphics[width=0.98\linewidth]{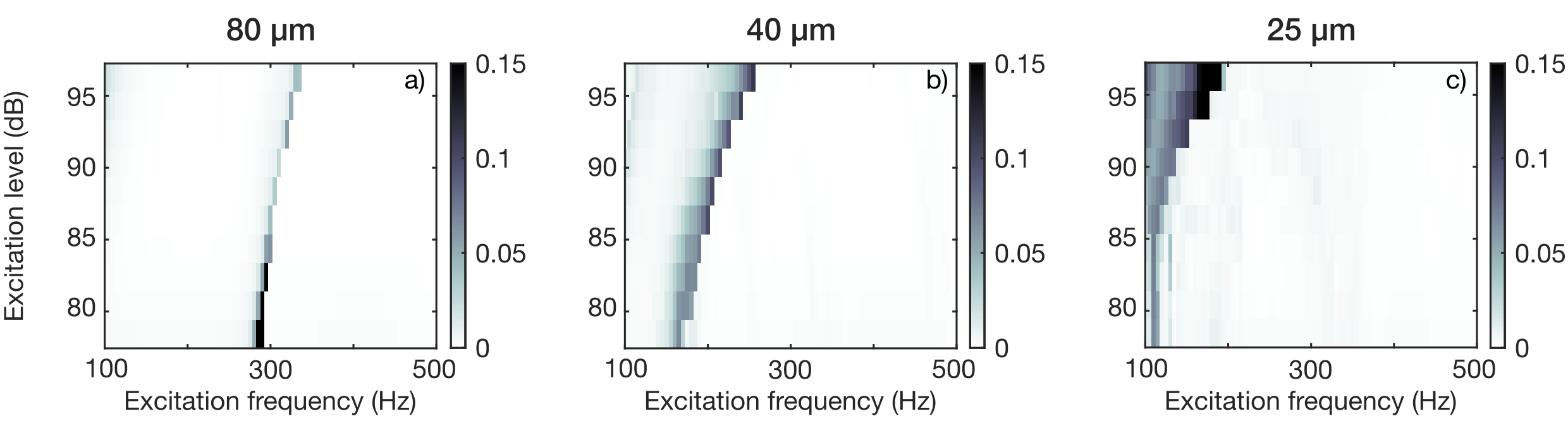}
    \caption{Transmitted power above the excitation frequency normalized by the total transmitted power, plotted against the level and frequency of excitation.}
    \label{fig:Paramac}
\end{figure*}

\section{Acoustic frequency-up conversion}

The maps in Figure \ref{fig:Paramac} illustrate the proportion of the transmitted acoustic power at frequencies higher than the excitation frequency. This indicator quantifies the acoustic frequency-up conversion efficiency, the main goal of the present study. These maps reveal a similar overall trend compared to the vibration data displayed in Fig.~\ref{fig:ParamVib}, with the highest conversion occurring at the resonance frequency of each plate. For the 80 $\mu$m plate, Fig.~\ref{fig:ParamVib}(a), the conversion is the largest at the lowest excitation amplitudes, where 20 percent of the power is located above the excitation frequency, and it decreases to 5 percent as the amplitude increases. The 40 $\mu$m plate has a consistent conversion rate of 10 percent around its resonance frequency, while the 25 $\mu$m plate reaches a maximum conversion of 30 percent around 200 Hz and at the maximum excitation level.

Interestingly, the maxima of acoustic frequency-up conversion correspond to the regimes where the dominant frequency of vibration is the fundamental, rather than the regimes where the harmonics are dominant. Surprisingly, it means that the regimes with maximum vibratory frequency-up conversion do not produce the most acoustic frequency-up conversion. Instead, it appears that the acoustic frequency up conversion is tied with the regimes that exhibit the highest amplitude of vibration at the frequency of forcing. This observation is consistent with the literature, which indicates that the acoustic radiation of a clamped plate is proportional to its surface velocity \cite{putra_sound_2010}. Consequently the maximum of velocity coincides with the maximum of radiation. Although this vibratory regime does not exhibit the most efficient vibratory frequency up conversion, it is the most acoustically efficient, and thus leads to the maximum acoustic frequency-up conversion.

\begin{figure*}[ht!]
    \centering
    \includegraphics[width=0.95\linewidth]{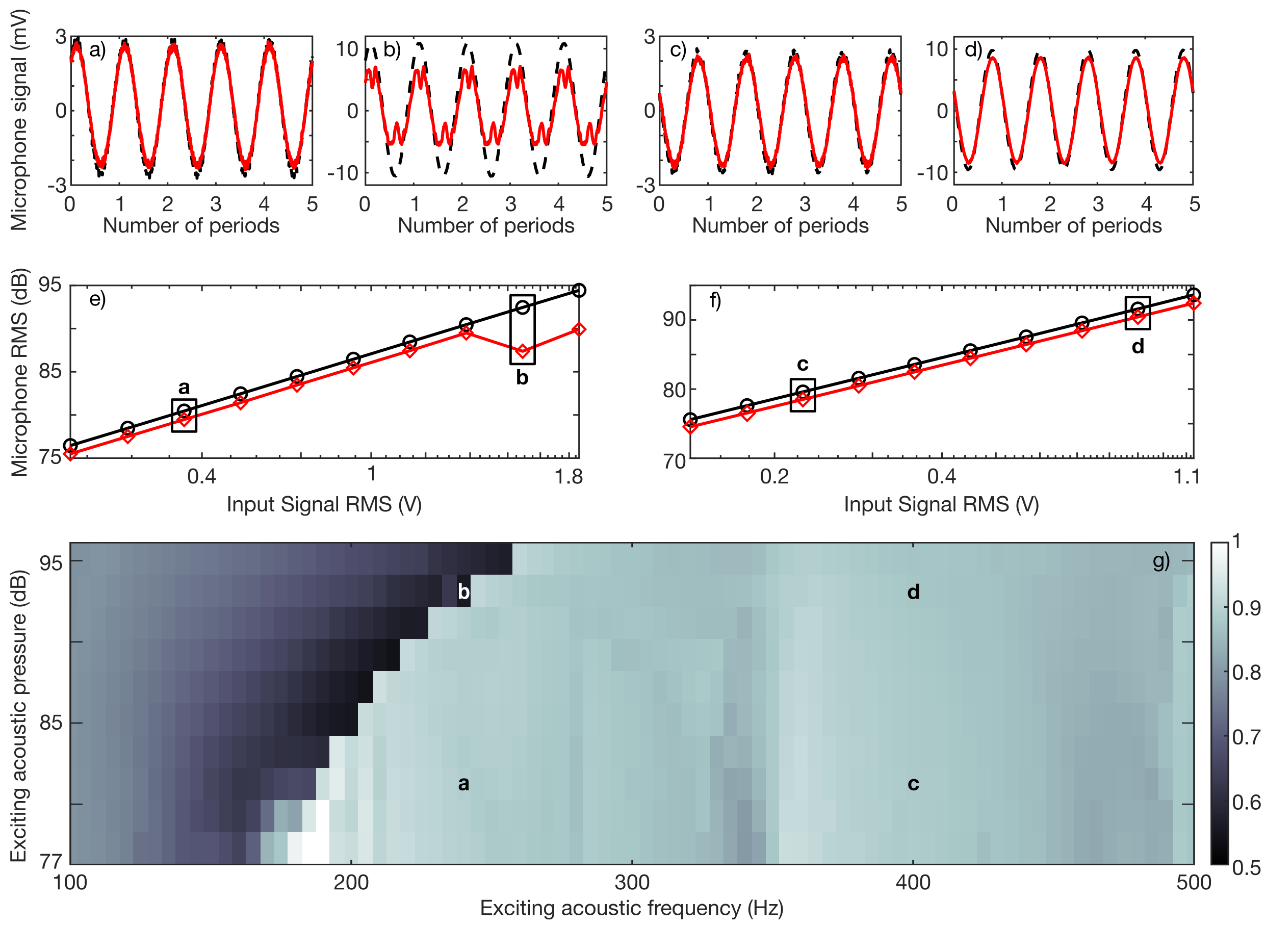}
    \caption{a. Microphonic signal of a 240 Hz sine at 81 dB (dotted black) compared to the signal transmitted (red) by the 40 $\mu$m plate under similar excitation. b. Microphonic signal of a 240 Hz sine at 93 dB (dotted black) compared to the signal transmitted (red) by the 40 $\mu$m plate under similar excitation. c. Microphonic signal of a 400 Hz sine at 81 dB (dotted black) compared to the signal transmitted (red) by the 40 $\mu$m plate under similar excitation. d. Microphonic signal of a 400 Hz sine at 93 dB (dotted black) compared to the signal transmitted (red) by the 40 $\mu$m plate under similar excitation. e. Root mean square of the microphonic signals at 125 from 74 to 92 dB captured without the nonlinear element (black), transmitted by the nonlinear element (red). f. Root mean square of the microphonic signals at 255 Hz from 74 to 92 dB captured without the nonlinear element (black), transmitted by the nonlinear element (red). g. Root mean square map of the signal transmitted by the nonlinear element normalized by the root mean square of the acoustic signal measured at the same point in absence of the nonlinear element. Letters in e-g corresponds to the time signals a-c.}
    \label{fig:RMSsat}
\end{figure*}

In addition to the frequency conversion effect, the resonance of the plate also has an effect on the amplitude of transmitted sound. Figure \ref{fig:RMSsat}a-d compares the transmitted sound in the presence of the 40 $\mu$m plate (in red), with the case where no plate is present (in black), at different frequencies and amplitudes of excitation. The behavior is similar between the cases with or without the plate, with the exception of Fig.~\ref{fig:RMSsat}.b. For this specific excitation, that corresponds to the resonance, the plate significantly lowers the amplitude of the transmitted acoustic signal.

This effect can be characterized further by comparing the RMS (Root Mean Square) levels of the acoustic signals for different excitations. It is computed by using the formula
\begin{equation}
    RMS=\sqrt{\frac{1}{n}\sum_i x_i^2},
\end{equation}
The map in Fig.~\ref{fig:RMSsat}.g corresponds to the ratio of the RMS level between the case with the plate and the case without. A tongue of minima is observable for an excitation frequency located between 160 Hz and 255 Hz, corresponding to the maxima of the vibration. Notably, this dynamics is localized at the resonance of the plate, but appears to be broadband at lower frequencies, although less efficient. Outside of the resonance, the ratio of RMS levels is relatively constant, as the plate acts mostly as a rigid boundary than a resonant object. 

This section highlights the efficiency of thin plates in the frequency up-conversion of monochromatic acoustic waves, both in vibration and in the radiated sound. Thickness was identified as a key physical parameter driving the dynamics of these plates, with thicker plates tending to exhibit a more linear behavior and resonance frequencies located higher in the frequency spectrum, while thinner plates are associated with lower resonance frequencies and a more nonlinear behavior. The role of static tension, as a factor that can modify the resonance frequency was discussed, potentially allowing us to fine-tune the frequency range of the effect we are studying. Lastly, we demonstrated that these plates can act as acoustic dampers, reducing transmitted sound by up to 6 dB in the resonant regime. In the following section, we explore the possibility of using a combination of plates to obtain effects over a wider range of frequencies.

\section{Panels of plates}

The previous section demonstrated the efficiency of the studied objects to convert a portion of the acoustic power towards higher frequencies through the effect of nonlinear vibration. This effect is localized, in both amplitude and frequency, around the first resonance frequency of the plate. To achieve a broadband frequency-up conversion effect, it is intuitive to combine different plates with different physical parameters.

\begin{figure*}[ht!]
    \centering
    \includegraphics[width=0.7\linewidth]{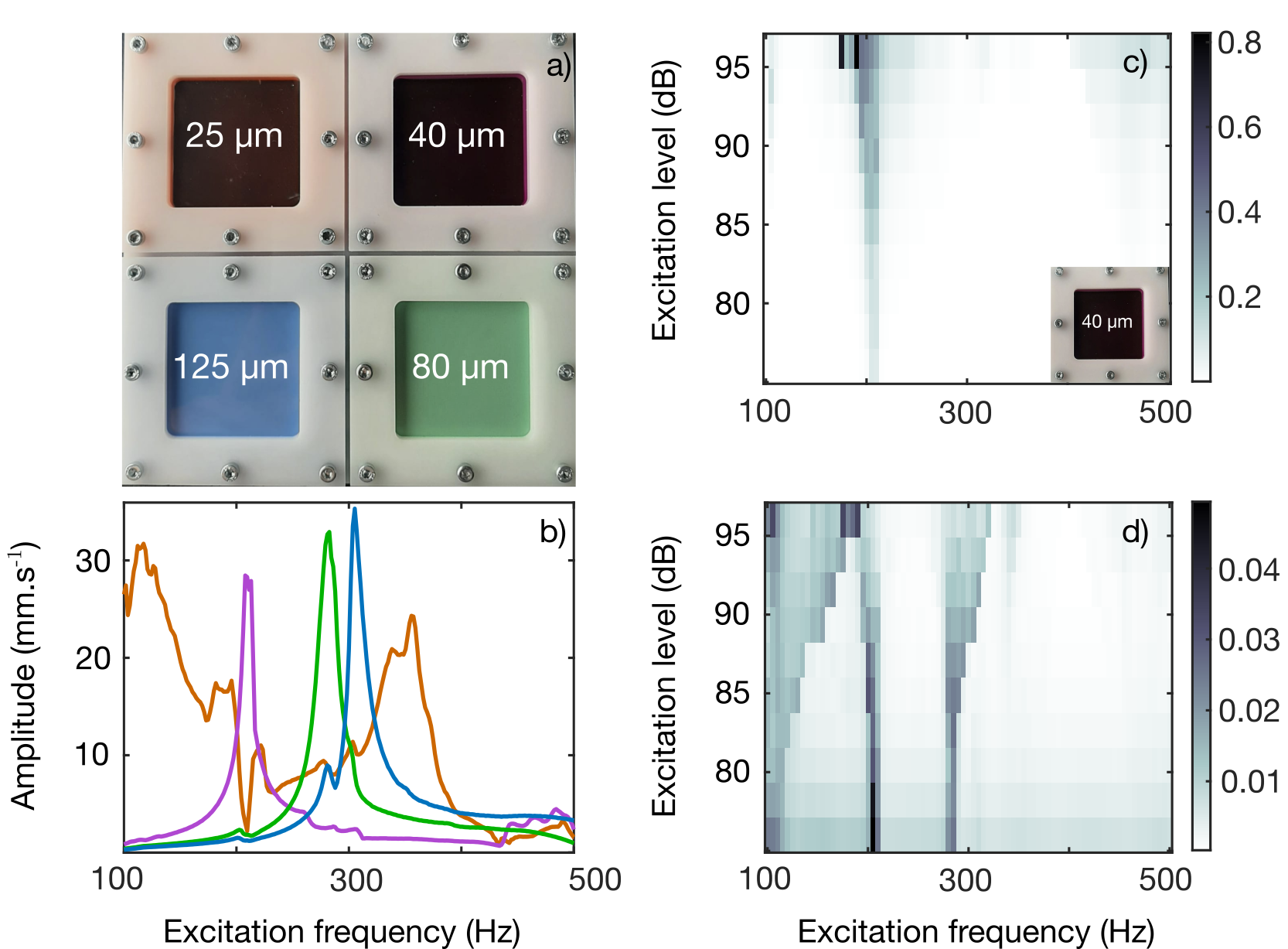}
    \caption{Experimental results for a panel made of 4 plates of different thicknesses". a) Picture of the panel, the thickness of each plate is written on the corresponding plate. b) Amplitude of vibration at the fundamental frequency measured near the center of each plate, 25 $\mu$m (orange), 40 $\mu$m (purple), 80 $\mu$m (green) and 125 micron thick (blue), in response to a swept sine excitation from 100 to 500 Hz at approximately 80 dB. c) Power Spectral Density map of the vibration of the 40$\mu$m thick sheet mounted in the panel of 4 plates. d) Proportion of transmitted acoustic power above the excitation frequency over the power of the total transmitted signal.}
    \label{fig:Fulfig7}
\end{figure*}

The first proposed design consists of a panel made up of four plates having different thicknesses (125, 80, 40, and 25 $\mu$m), as depicted in Fig.~\ref{fig:Fulfig7}. Preliminary tests on this panel configuration showed that the acoustic source directivity negatively affected the results, as the plates were not excited with a similar pressure compared to the single plate case. To address the directivity issue and enable simultaneous excitation of the different plates, a 3D-printed horn was designed and adapted to the sound source.

The measurement setup remains unchanged, meaning that only one plate can be monitored by the laser vibrometer during each experiment. To gather information about the vibration of each plate without repeating the experiment four times, a test experiment is conducted. In this experiment, the panel is excited with a swept sine wave ranging from 100 Hz to 500 Hz over a 40 second period. This experiment is repeated four times, with the laser vibrometer directed at a different plate in each iteration.

Figure \ref{fig:Fulfig7}.b shows the amplitude of the power spectrum at the frequency of excitation, i.e. it is a representation of the response that only contains the fundamental of the signal, without any harmonics. In Figure \ref{fig:Fulfig7}.b), five main peaks can be identified. The first four peaks correspond to the first resonance of each individual plate: 25, 40, 80, and 125 $\mu$m, in order of increasing frequency. The fifth peak corresponds to a secondary resonance of the 25 $\mu$m plate, most likely the second mode, although a single point of measurement for the vibration cannot give a definitive answer.

This increase in the frequency of resonance with the thickness of the observed plate is consistent with the results presented in section III, notably in Fig.~\ref{fig:ParamVib}, and indicates a plate-like behavior. In addition, a weak solid coupling between the different plates can be observed, as the main peaks in amplitude coincides with small increase in amplitude of the vibration of the other plates, observable in Fig.~\ref{fig:Fulfig7}.b at excitation frequencies of 203 Hz, 277 and 305 Hz. This is due to the clamping of the plates within the same panel, which leads to an elastic coupling between each plate.

Figure \ref{fig:Fulfig7}.c shows the vibration map of the 40 $\mu$m plate clamped into the panel, excited by a single frequency at different amplitude as described in the first section. It shows a single tongue of resonance, located between 255 Hz at 77 dB and 180 Hz at 95 dB, which corresponds to the peak in Fig.~\ref{fig:Fulfig7}.c. It is notable that the behavior of this tongue is different from the one studied in the first section. The frequency at which the tongue is observed is similar, at 200 Hz, but the tongue in Fig.~\ref{fig:ParamVib}.b clearly bends towards high frequencies when the amplitude increases, while the tongue showed in Fig.~\ref{fig:Fulfig7}.d bends towards the low frequencies under similar excitation. This result is surprising, but can probably be attributed to a difference in the clamping of this specific plate, as the clamping is the main issue in the reproducibility of the observed results so far. While the vibration measurements only gives information on a single plate, the radiated acoustic field should contain components from each independent plate. It is observed in Fig.~\ref{fig:Fulfig7}.d that there are four distinct tongues of frequency-up conversion, each one corresponding to the first resonance of a plate. This also leads to a broadband frequency conversion between 100 and 200 Hz at 95 dB of excitation,  with a local maximum where two resonance tongues cross at 190 Hz. The tongue from 285 Hz to 315 Hz is linked to the resonance of the 80 $\mu$m thick plate and is located in the same frequency/amplitude range as the single 80 $\mu$m thick plate, as observed in Fig.~\ref{fig:Paramac}.a for the case of a single plate.

The last peak at 355 Hz is much less visible and corresponds to the 125 $\mu$m thick plate. This plate was not tested in the previous section, as it was deemed too thick to exhibit interesting effect, but was chosen for the panel configuration in order to reach broadband properties. Its higher thickness leads to a smaller effect on the acoustic distortion, since the nonlinear behavior of these plates is tied to the ratio between their displacement and their thickness. This confirms the hypothesis that a thickness of 120 $\mu$m is too high for this plate to exhibit efficient frequency-up conversion.

The proportion of frequency-up conversion is below what was measured for single plates, reaching five instead of fifteen percent, which might be explained by the change in the directivity of the sound source. 

\begin{figure*}[ht!]
    \centering
    \includegraphics[width=0.7\linewidth]{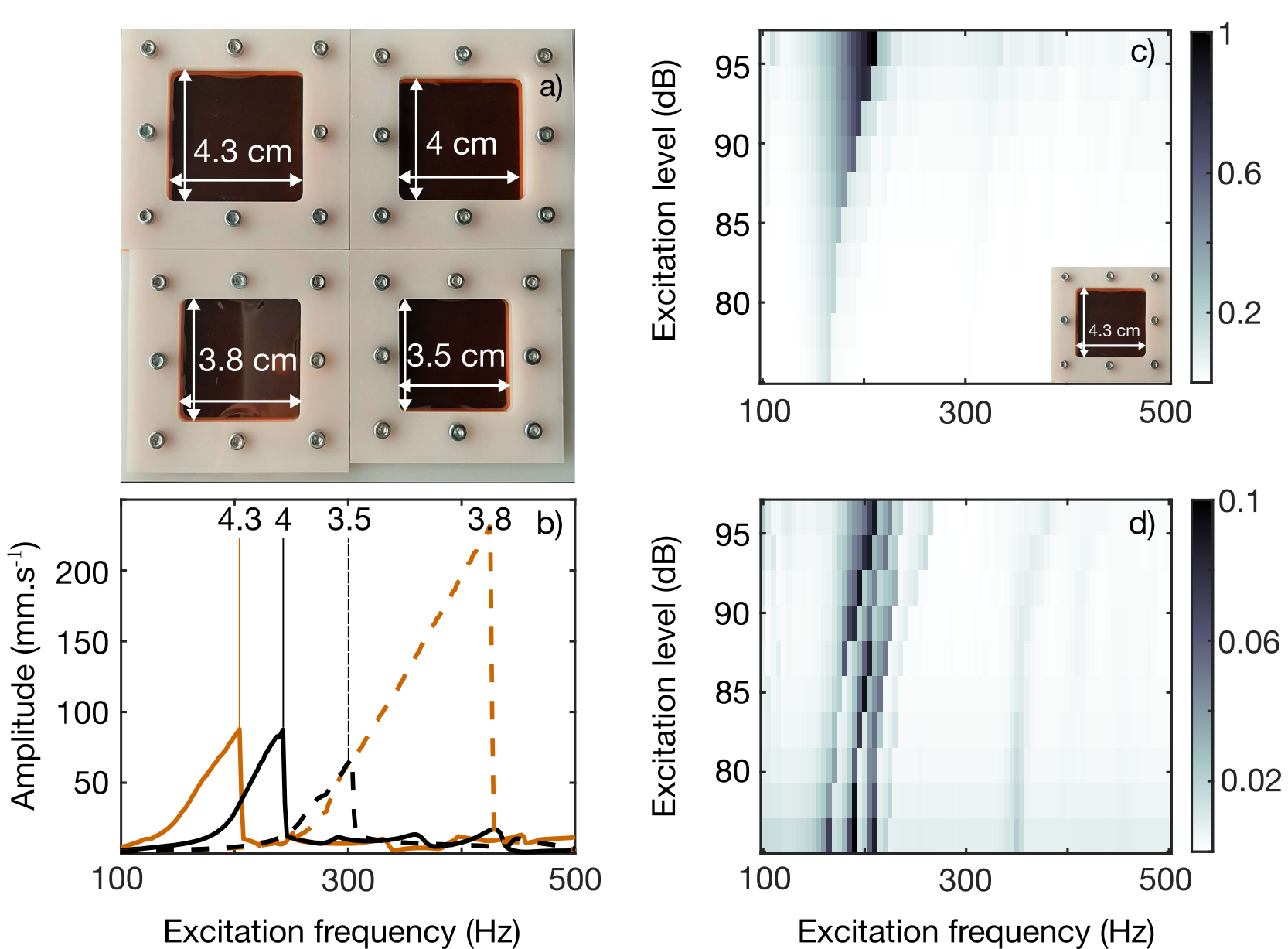}
    \caption{Experimental results for a panel made of 4 plates of different sizes and the same thickness. a) Picture of the panel, the side length of each plate is written on the corresponding plate. b) Amplitude of vibration at the fundamental frequency measured near the center of each plate, 4.3 cm (orange), 4 cm (black), 3.8 cm (dashed orange) and 3.5 micron thick (dashed black), in response to a swept sine excitation from 100 to 500 Hz at approximately 80 dB. c) Power Spectral Density map of the vibration of the 40$\mu$m thick sheet mounted in the panel of 4 plates. d) Proportion of transmitted acoustic power above the excitation frequency over the power of the total transmitted signal. }
    \label{fig:Fulfig8}
\end{figure*}

Overall, this panel configuration tends to show that broadband frequency-up conversion can be achieved by assembling several nonlinear plates. While the thickness is the main parameter studied so far, other dimensions of the plates can be changed in order to tune their response. Following this idea, the second panel tested consists of plates of the same thickness, 25$\mu$m, but of different side length, namely 4.3, 4, 3.8 and 3.5 cm, as displayed in Fig.~\ref{fig:Fulfig8}.a. A slightly different method was used to make this panel compared to the panel in Fig.~\ref{fig:Fulfig7}.a. The results in Fig.~\ref{fig:Fulfig7}.b show a coupling between each plate through the elastic frame, which complicates the analysis of the results. This second panel consists of separately clamped plates that can then be assembled in a panel, which allows for the study of their individual behavior, better decoupled from the panel.

Figure \ref{fig:Fulfig8}.b shows the response at the fundamental resonance frequency of each different plate before clamping them in the panel. The main observation is that every plate presents a very similar qualitative response, similar to a saw tooth. This shape is characteristic of the hardening behavior of the first mode of the plate. The increase in amplitude of vibration shifts the resonance frequency toward higher frequencies, as the excitation signal increases in frequency, leading to an increase in the vibration amplitude. At a specific threshold in frequency, where the excitation frequency exceeds the resonance frequency, causing the system to fall out of resonance, leading to this sawtooth shape. Another observation is that the frequency distribution of the different peaks in Fig.\ref{fig:Fulfig8}.b is not the one expected. Plate behavior would lead to an increase in the frequency of the first mode that would depend on the side length of the plate, such that the lowest frequency would correspond to the 4.3$\times$4.3 cm plate (in orange) and the highest would correspond to the 3.5$\times$3.5 cm plate (in dashed black). However, the experimental results contradict this initial assumption. The proposed explanation is as follows: the static tension introduced by the clamping is unknown in the experiment, and hard to characterize beforehand. It has a great impact on the eigenfrequencies of the plate, especially for the thinnest plates. An improved version of the MAGPIE framework \cite{hamilton_magpie_2024} (an eigenmode estimator for square plates) is used to compute the first eigenfrequency of the 4.3$\times$4.3cm plate of 25$\mu$m thickness. The numerical results show that an increase in the tension from 0 to 4 N/m lead to a shift from 105 Hz to 216 Hz in the frequency of the first mode, which would explain the discrepancies between two measurements of plates of the same thickness and size.

Figure \ref{fig:Fulfig8}.c displays the vibration map of the 4.3$\times$4.3 cm plate assembled on the panel. It shows a tongue of resonance similar to the one observed in precedent tests on the same plate, although shifted up in frequency compared to the tests on a similar plate displayed in Fig.~\ref{fig:ParamVib}.c. This tongue corresponds to the leftmost one observable in Fig.~\ref{fig:Fulfig8}.d, which depicts the map of the acoustic frequency-up conversion of the full panel. It shows three different tongues of maxima located at 165, 185 and 215 Hz at their lowest amplitude. These three tongues corresponds to the first three peaks in Fig.~\ref{fig:Fulfig8}.b and convert from five to ten percent of the incident power of the wave to higher frequencies in a band of 50 Hz at maximum excitation amplitude. At 350 Hz, another tongue of maxima corresponds the resonance of the 3.8$\times$3.8 cm plate, which appears to be less efficient in frequency-up conversion, reaching a maximum of three percent.

This configuration leads to promising, though difficult to predict, results in terms of acoustic frequency-up conversion. While variations in plate thickness in the first panel produces the expected effects, changes in plate size leads to high uncertainty on the static tension introduced by the clamping. This makes it difficult to predict the plate first eigenfrequency, and thus what frequency will lead to the maximum frequency-up conversion. Despite these uncertainties, small differences in eigenfrequencies between plates result in a quasi-broadband effect. This suggests a valuable design principle for future systems: intentionally introducing slight variations in plate properties, in order to recover a broadband frequency conversion. Overall, the two configurations presented in this section demonstrate great capabilities for acoustic frequency-up conversion at different frequencies.
\section{Conclusion}

This study demonstrates that thin, clamped plates subjected to acoustic excitation in air can exhibit strongly nonlinear vibrational behavior, leading to the conversion of a portion of the incident acoustic energy from low to high frequencies. Through a combination of laser vibrometry and microphonic measurements, the link between the first resonance of the plate and the acoustic frequency-up conversion effect was observed. A parametric study on the thickness of the plate highlights its importance in the localization of the regimes of interest, but also reveals that the clamping plays an important role in the dynamics by inducing a static tension that shifts the resonance frequencies.
Moreover, the exploration of more complex geometries, involving multiple plates of different thicknesses and sizes, suggests that the acoustic frequency-up conversion effect could be tuned or enhanced by design. The main obstacle to such design seems to reside in the extreme sensitivity of the static tension to the clamping, leading to reproducibility issues and uncertainty in the frequencies of maximum efficiency of the plates.





\end{document}